\documentclass[preprint,aps,showkeys,draft]{revtex4}

\usepackage{amsfonts,amsmath,amssymb,amsthm}
\usepackage{bbm,bm}
\usepackage{graphics,graphicx}
\usepackage{latexsym}
\usepackage{color} %\usepackage{times}

\newcommand{\ra}{\rangle}
\newcommand{\la}{\langle}

\newcommand{\e}{\mathrm{e}}
\newcommand{\g}{\gamma}
\newcommand{\x}{{\bm x}}
\newcommand{\y}{{\bm y}}

\newcommand{\s}[1]{\sqrt{#1}}
\newcommand{\p}{\prime}
\newcommand{\xx}{{\bm x}^\p}
\newcommand{\yy}{{\bm y}^\p}
\newcommand{\z}{{\bm s}}
\newcommand{\tr}{\rm Tr}
\newcommand{\ket}[1]{\vert #1 \ra}

\renewcommand{\l}{\lambda}
\renewcommand{\b}{\bf b}

\renewcommand{\o}{\otimes}

\begin{document}

\title{Completely mixed state is a critical point for three-qubit entanglement}

 \author{Sayatnova Tamaryan}
 \affiliation{Department of Theoretical Physics, A. Alikhanyan National Laboratory, Yerevan, Armenia\\
 {\rm Email: sayat@mail.yerphi.am, Phone: +[374] 99 485 402 }}

 \begin{abstract}
Pure three-qubit states have five algebraically independent and one algebraically dependent polynomial invariants under local unitary transformations and an arbitrary entanglement measure is a function of these six invariants. It is shown that if the reduced density operator of a some qubit is a multiple of the unit operator, than the geometric entanglement measure of the pure three-qubit state is absolutely independent of the polynomial invariants and is a constant for such tripartite states. Hence a one-particle completely mixed state is a critical point for the geometric measure of entanglement.
 \end{abstract}

%\pacs{03.67.Mn, 03.67.Bg, 74.40.Kb}

\keywords{entanglement measures, quantum critical phenomena, teleportation, geometric measure}

\maketitle

\section{Introduction}

The concept of quantum entanglement has played an important role in the development of modern quantum
physics. The entanglement theory has its roots in the key discoveries: quantum cryptography~\cite{ek-91,exp-cr}, dense coding~\cite{dense-92,exp-den}, perfect
teleportation~\cite{tele-93,exp-tel} and quantum factoring algorithm~\cite{shor-94,shor-exp}. These
effects are based on entanglement and all of them have been demonstrated in pioneering experiments.

However, the phrase "based on entanglement" does not reveal the essence of those effects. For instance,
one would like to know what are the states that can be used as a quantum channel for perfect
teleportation and dense coding? And why those states are capable, while others are not?  In the case of
bipartite systems these questions have a concise answer, namely, the maximally entangled states can
perform the task and others cannot \cite{fidel}. But the situation is changed drastically in a multipartite setting. For instance, in the case of three-qubit systems among the states applicable for teleportation are:  the Greenberger-Horne-Zeilinger(GHZ) state~\cite{ghz-tel} which is maximally entangled and a W-class state~\cite{w-tele} which is not maximally entangled since the only maximally entangled state within W-class states~\cite{w} is the W state~\cite{maximal}. Therefore in the case of multipartite systems the property of being maximally entangled is unrelated to the ability to perform a certain task.

In the meantime the aforementioned three-qubit states, as well as two-qubit maximally entangled states,
that are quantum channels for perfect teleportation have two common properties. First, in each state the
reduced density operator of a some qubit is a scalar multiple of the unit operator. Second, all those
state have the same maximal product overlap~\cite{wei-03,shared,hyper}. Recall that the maximal product overlap $g$ of a pure state $\ket{\psi}$ is given by
\begin{equation}\label{1.mpo}
   g(\psi)=\sup_{q_A,q_B,q_C}|\la\psi|q_Aq_Bq_C\ra|,
\end{equation}
where the maximization runs over all product states and the normalization of the local
states $\ket{q_M}(M=A,B,C)$ is understood.

Then one makes a guess that these two properties are interrelated and the reasons of the interrelation should be analyzed. Moreover, if a pure state has a bipartite entanglement that does not depend on the bipartition and is maximal for all possible bipartitions then it should have both these properties~\cite{bipart}.

In this article we consider pure three-qubit states and prove that if the density matrix of one qubit is
a completely mixed state, then the maximal product overlap $g$ of the state is $1/\s{2}$.  The basic
point of the proof is the existence of the specific point in the space of entanglement
parameters~\cite{param}. In the case of pure three-qubit states there are five algebraically independent
and one algebraically dependent polynomial invariants under local unitary transformations~\cite{inv}.
They can be regarded as coordinates on the space of entanglement types and thus any entanglement measure
should be a function of these invariants. The three polynomials quartic in state function are the
squares of the lengths of the Bloch vectors and they play a crucial role. If one of these three quartic
invariants vanishes, then the sextic polynomial vanishes too. Given that all invariants are independent,
the vanishing of the sextic polynomial indicates the existence of a specific point. Hereafter these
points are referred to as critical points.

The term critical point has a mathematical and a physical justifications. The mathematical justification
is the following. The geometric entanglement measure of quadrilateral three-qubit~\cite{shared} and
general W~\cite{dual} states have been computed analytically and the answers show the the gradient of
the measure has a jump at these points. Hence they are critical points. The physical justification is
that at the edge of the region of possible values of a quartic polynomial the entanglement of the state
is absolutely independent of the remaining five invariants and then the state acquires an ability to be
a quantum channel for the perfect teleportation and dense coding.

This article is organized as follows. In Sec. II, we review local unitary invariants of pure three-qubit
states. In Sec. III, we formulate and prove the main theorem.  In Sec. IV, we show the for four- and higher-qubit states the theorem ceases to be true. In Sec. V, we discuss our results.

\section{Local unitary invariants of pure three-qubit states}

In this section we briefly review local unitary invariants of pure three-qubit states~\cite{inv} and
slightly modify original definitions for convenience. Consider a pure state $\psi$ of the three qubits
A, B and C.

There is one independent invariant quadratic in $\psi$  which is just the norm of the three-party state
and therefore has no physical significance. We set it equal to 1.

There are three independent quartic invariants, namely $\tr(\rho_A^2), \tr(\rho_B^2)$ and
$\tr(\rho_C^2)$, where $\rho_A,\rho_B$ and $\rho_C$ are the one-particle density operators of the qubits
A, B and C, respectively. We redefine these invariants as follows
 \begin{equation}\label{2.inv123}
   b_A^2=2\tr(\rho_A^2)-(\tr\rho_A)^2,\; b_B^2=2\tr(\rho_B^2)-(\tr\rho_B)^2,\;
   b_C^2=2\tr(\rho_C^2)-(\tr\rho_C)^2.
 \end{equation}
Of course, the substraction of the constant $(\tr\rho_A)^2=(\tr\rho_B)^2=(\tr\rho_C)^2=1$
does not change anything and the new invariants are independent too. The advantage of the redefinition
is that the new invariants $b_A,\;b_B$ and $b_C$ are the lengths of the Bloch vectors ${\b}_A,\;{\b}_B$
and ${\b}_C$ of the qubits A, B and C, respectively. Therefore $b_M=0\;(M=A,B,C)$ means that $\rho_M$ is
a completely mixed state and $b_M=1$ means that $\rho_M$ is a pure state.

There is one sextic invariant given by
 \begin{equation}\label{2.inv4}
   t=3\tr[\rho_{AB}(\rho_A\o\rho_B)]-\tr(\rho_A^3)-\tr(\rho_B^3)-\frac{1}{4}(\tr\rho_c)^3.
 \end{equation}
 Again our definition slightly differs from the original one since we subtracted the last
term(equal to 1/4) that does not exist in the original definition. The meaning of the substraction of
the last term is the following. Let us define a bipartite correlation matrix $G$ by formula
\begin{equation}\label{2.def} G_{ij}=\tr(\rho_{AB}\sigma_i\o\sigma_j),
\end{equation}
where $\sigma_i$'s are Pauli matrices. Then $t$ is expressed via the Bloch vectors and the correlation matrix as follows
\begin{equation}\label{2.tg}
  t=\frac{3}{4}\;{\b}_A\cdot(G{\b}_B),
\end{equation}
which does not contain an additional term 1/4.

The invariant of degree 8 is (up to a numerical factor) the square of the three-tangle
$\tau$~\cite{coff}. We use it as is.

The last invariant $i$ is discrete in the sense that it can have at most two different values when the
other invariants are fixed~\cite{grov}. It appears as follows. The polynomial invariants constructed in
Ref.~\cite{inv} do not distinguish a state and its complex conjugate and therefore one additional
invariant is needed to specify uniquely a pure three qubit state. However, all amplitudes of a
three-qubit pure state can be chosen positive when a Bloch vector vanishes as will be shown below and
then the additional invariant does not play any role here. This point will be explained more clearly
below.

We listed all independent invariants that are coordinates on the space of orbits of the group of local
transformations. An arbitrary entanglement measure, and among them the geometric measure, is a function
of these variables~\cite{inv,onish}.

\section{Main theorem}

It is curious that there are specific points in the space of entanglement parameters $(b_A,
b_B,b_C,t,\tau,i)$. The following theorem clarifies the specific points and the physical meaning of
those points.

{\bf Theorem.}
 $${\rm If}\;\; b_Ab_Bb_C=0,\;\; {\rm then}\;\; g^2=\frac{1}{2}.$$
The main objective of the article is this theorem. It states that if any of invariants  $b_A,\;b_B$ or
$b_C$ vanishes, then the geometric measure given by \begin{equation}\label{2.measure}
   E_g(\psi)=-2\ln g(\psi)
\end{equation} is absolutely independent of the remaining entanglement parameters and is a constant for these states.

To prove the theorem we use the generalized Schmidt decomposition that is closely related to the
geometric measure~\cite{hig} and hence it is more appropriate to our analyze here. It states that a pure
state $\ket{\psi}$   has the canonical form~\cite{hig,maximal}
\begin{equation}\label{2.psi} \ket{\psi}
= a\ket{011}+b\ket{101}+c\ket{110}+d\ket{000}+\e^{i\g} h\ket{111},
\end{equation}
where labels within kets refer to the qubits A, B and C in that order. All the coefficients $a,b,c,d,h$ in \eqref{2.psi} are positive and the gauge phase $\g$ ranges from $-\pi/2$ to $\pi/2$.

We need to compute explicitly the Bloch vectors and correlation matrix for the following analysis. The
computation is straightforward and yields
\begin{eqnarray}\label{2.bloch} % \nonumber to remove
 {\b}_A &=&\left(2ha\cos\g,\;2ha\sin\g,\;d^2+a^2-b^2-c^2-h^2\right), \\ \nonumber
 {\b}_B &=& \left(2hb\cos\g,\;2hb\sin\g,\;d^2+b^2-a^2-c^2-h^2\right), \\ \nonumber
 {\b}_C &=& \left(2hc\cos\g,\;2hc\sin\g,\;d^2+c^2-b^2-a^2-h^2\right)
 \end{eqnarray}
and
\begin{equation}\label{2.cor}
 G=
\begin{pmatrix}
 2ab+2cd & 0 & -2ha\cos\g \\
 0 & 2ab-2cd & -2ha\sin\g \\
 -2hb\cos\g & -2hb\sin\g & d^2-a^2-b^2+c^2+h^2
 \end{pmatrix}.
\end{equation}

Using these expression one can calculate five polynomial invariants and see that they are all even
functions on $\g$. For instance, the three tangle is
\begin{equation}\label{2.tau}
   \tau=4d\s{(dh^2-4abc)^2+16abcdh^2\cos^2\g}.
\end{equation}
Next one can inverse the obtained relations and express state parameters $a,b,c,d,h$ and
$\g$ via polynomials $|\psi|^2,b_A,b_B,b_C,t$ and $\tau$(see equation (3.4) and comment below in
Ref.\cite{grov}). But the inverse solution is not unique since both signs + or - are equally good for
the gauge phase $\g$. Right here it appears a necessity to introduce  the additional invariant $i$ that
distinguishes positive and negative values of the gauge phase $\g$.

But to prove the theorem we do not need $i$ at all. Indeed, from ${\b}_C=0$ it follows that either $h=0$
or $c=0$. Now $\psi$ is a linear combination of four orthogonal product states and then the phase can be
eliminated by appropriate local unitary transformations. Similarly, no additional invariant is needed if
either ${\b}_A=0$ or ${\b}_B=0$.

\subsection{Stationarity equations.}  The maximal product overlap $g(\psi)$ of a pure state
$\ket{\psi}$ can be expressed via the Bloch vectors and correlation matrix as follows~\cite{shared}
\begin{equation}\label{2.mpomix}
   g=\sup_{x^2=y^2=1}\frac{1}{4}\left[1+\x\cdot{\b}_A + \y\cdot{\b}_B + \x\cdot (G\y)\right],
\end{equation}
where maximization runs over all unit vectors $\x$ and $\y$. Note that $\x$ and $\y$ are
the Bloch vectors of the local states $\ket{q_A}$ and $\ket{q_B}$, respectively.

By introducing Lagrange multipliers $\l_1$ and $\l_2$ that enforce unit vectors $\x$ and $\y$ one
obtains the following stationarity equations:
\begin{subequations}\label{2.stat}
\begin{equation}\label{2.xstat}
 G\y+{\b}_A=\l_1\x,
\end{equation} \begin{equation}\label{2.ystat}
 G^T\x+ {\b}_B=\l_2\y.
\end{equation} \end{subequations}
Unknown Lagrange multipliers $\l_1$ and $\l_2$ are defined by the conditions
\begin{equation}\label{2.cond}
    |\x|^2=1,\quad |\y|^2=1.
\end{equation}
These conditions are a pair of algebraic equations of degree six in $\l_1$ and $\l_2$.
\'Evariste Galois's theory states that there is no general answer in terms of radicals and then the only
possibility to find roots is the factorization of algebraic equation. The factorization of generic
polynomials is impossible and therefore one cannot expect a simple closed form solution for arbitrary
three-qubit states. Similar results have been obtained in Ref.\cite{hyper}, where the authors derived a
polynomial expression of degree 12 in the coefficients of a three-qubit pure state, the roots of which
include the maximal product overlap of the state.

Our main tasks are to solve the couple of equations \eqref{2.stat} when the length of a some Bloch
vector vanishes and show that $g^2=1/2$ in this case. It suffices to consider only the case $b_C=0$
since the analysis of the remaining cases is similar. Furthermore, from $b_C=0$ it follows that either
$h=0$ or $c=0$. These two cases will be analyzed separately in the next sections.

\subsection{Quadrilateral states}

In this section we consider the case $h=0$. Now the correlation matrix $G$ is diagonal and this fact
essentially simplifies the stationarity equations.

The state \eqref{2.psi} is equivalent, up to local unitary transformations, to the state
\begin{equation}\label{3.psi}
\ket{\psi^\p }= a\ket{100}+b\ket{010}+c\ket{001}+d\ket{111}.
\end{equation}
Fortunately this state is analyzed in detail in Ref.\cite{shared}, where it is shown that
an arbitrary entanglement measure is a fully symmetric function on state parameters $a,b,c,d$ in this
case. In particular, the three-tangle~\cite{coff} of the state \eqref{3.psi} is $\tau=16abcd$ and thus
$\psi^\p$ is the GHZ state when $a=b=c=d$ and is a W state when $abcd=0$~\cite{w}.

The maximal product overlap of $\psi^\p$ is(up to factor 2) the circumradius of the cyclic quadrangle
with the sides $a,b,c,d$ and therefore this type of states can be categorized as quadrilateral states.
The nearest product states of these states can be computed analytically by solving the stationarity
equations \eqref{2.stat}. In particular, the local constituents of the nearest product state of
$\psi^\p$ are~\cite{shared}
\begin{eqnarray}\label{3.nearh}
 \ket{q_A} &=& \frac{\s{r_ar_d}\,\ket{0_A}+\s{r_br_c}\,\ket{1_A}}{4S\s{ad+bc}}, \\ \nonumber
 \ket{q_B} &=& \frac{\s{r_br_d}\,\ket{0_B}+\s{r_ar_c}\,\ket{1_B}}{4S\s{bd+ac}}, \\ \nonumber
 \ket{q_C} &=& \frac{\s{r_cr_d}\,\ket{0_C}+\s{r_ar_b}\,\ket{1_C}}{4S\s{cd+ab}},
\end{eqnarray}
where \begin{eqnarray*}\label{3.rih}
  r_a &=& a(b^2+c^2+d^2-a^2)+2bcd, \\
  r_b &=& b(a^2+c^2+d^2-b^2)+2acd, \\
  r_c &=& c(b^2+a^2+d^2-c^2)+2abd, \\
  r_d &=& d(b^2+c^2+a^2-d^2)+2abc
\end{eqnarray*}
and $S$ is the area of the cyclic quadrangle $a,b,c,d$.

From $b_C=0$ and $h=0$ it follows that \begin{equation}\label{3.equal} c^2+d^2 = a^2+b^2. \end{equation}
Note that owing to this condition the correlation matrix $G$ acquires a zero eigenvalue. But we were not
forced to use it since the general solution for $h=0$ was already found in Ref.\cite{shared}.

The condition \eqref{3.equal} simplifies expressions \eqref{3.nearh} for the nearest product as follows
\begin{eqnarray}\label{3.nearsim}
 \ket{q_A} &=& \frac{\s{bc}\,\ket{0_A}+\s{ad}\,\ket{1_A}}{\s{ad+bc}}, \\ \nonumber
 \ket{q_B} &=& \frac{\s{ac}\,\ket{0_B}+\s{bd}\,\ket{1_B}}{\s{ac+bd}},\\ \nonumber
 \ket{q_C} &=& \frac{\s{dc}\,\ket{0_C}+\s{ab}\,\ket{1_C}}{\s{ab+cd}}.
 \end{eqnarray}
The substitution of these expressions into Eq.\eqref{1.mpo} gives
\begin{equation}\label{3.mpo}
    g = \frac{(c^2+d^2)ab + (a^2+b^2)cd}{\s{(ad+bc)(ac+bd)(ab+cd)}}.
\end{equation}
Now from the identity $$(ac+bd)(bc+ad)=(c^2+d^2)ab + (a^2+b^2)dc$$ and the normalization
condition it follows that
\begin{equation}\label{3.final}
   g^2=\frac{1}{2}.
\end{equation}

\subsection{The case $h\neq0$.}

In this section we consider the case $c=0$. Then the three-tangle is $\tau=d^2h^2/4$ and the state is:
the GHZ state when $d=h$ and $a=b=0$, a W state when $h=0$ and a biseparable state when $d=0$.

In general, if $h\neq0$ then the system of stationarity equations are unsolvable since Eq.\eqref{2.cond}
yield generic algebraic equations of degree six. Moreover, the method developed in Ref.\cite{hyper} also
results a  nonfactorizable characteristic polynomial of degree 12 for $h\neq0$. In this reason the
geometric measure of these states has not been investigated so far except particular
cases~\cite{symmet,ilum,bast,lin-1}.

Now we solve the stationarity equations explicitly when $h\neq0$ but $b_C=0$. Fortunately when $b_C=0$
the correlation matrix $G$ has a zero singular value and it is of crucial importance here. Owing to the
existence of the zero singular value the sextic equations \eqref{2.cond} can be factorized to the linear
and quadratic equations. However, we use a singular value decomposition~\cite{book} instead to get rid
of laborious algebra and find the answer as quick as possible.

\subsubsection{Singular value decomposition} The requirements $b_C=0$ and $c=0$ impose the following
condition
\begin{equation}\label{4.equal}
d^2=a^2+b^2+h^2.
\end{equation}
Furthermore, the gauge phase $\g$ can be eliminated by appropriate local unitary transformations and we set $\g=0$ for the simplicity. Then nonzero Bloch vectors are
\begin{equation}\label{4.bloch}
{\b}_A =b_A(\sin\alpha,\;0,\;\cos\alpha),\quad{\b}_B = b_B(\sin\beta,\;0,\;\cos\beta),
\end{equation}
where
$$b_A=2a\s{h^2+a^2},\quad b_B=2b\s{h^2+b^2},\quad \tan\alpha=\frac{h}{a},\quad \tan\beta=\frac{h}{b}.$$
The correlation matrix $G$ is given by the simplified formula
\begin{equation}\label{4.gmatr}
    G=
\begin{pmatrix}
 2ab & 0 & -2ha \\
 0 & 2ab & 0 \\
 -2hb & 0 & 2h^2
\end{pmatrix}.
\end{equation}
The singular value decomposition of $G$ is
\begin{equation}\label{4.sing}
   G=UDV^+,
\end{equation}
where
\begin{equation}\label{4.udv}
   U=
\begin{pmatrix}
 \cos\alpha & 0 & \sin\alpha \\
 0 & 1 & 0 \\
 -\sin\alpha & 0 & \cos\alpha
\end{pmatrix},\;\,
 D=
\begin{pmatrix}
 2\mu & 0 & 0 \\
 0 & 2ab & 0 \\
 0 & 0 & 0
\end{pmatrix},\;\,
 V=
\begin{pmatrix}
 \cos\beta & 0 & \sin\beta \\
 0 & 1 & 0 \\
 -\sin\beta & 0 & \cos\beta
\end{pmatrix} \end{equation}
and $$\mu=\s{(h^2+a^2)(h^2+b^2)}.$$

It is easy to see that
\begin{equation}\label{4.rebloch}
  {\b}_A=b_AU\z,\quad{\b}_B=b_BV\z,
\end{equation}
where $\z=(0,0,1)$ is the eigenvector of $D$ with zero eigenvalue, i.e
\begin{equation}\label{4.zero}
   D\z=0.
\end{equation}
Now we define new unit vectors $\xx$ and $\yy$ as follows
\begin{equation}\label{4.xyprim}
   \xx=U\x,\quad\yy=V\y.
\end{equation}
Substituting expressions \eqref{4.sing}, \eqref{4.rebloch} and \eqref{4.xyprim} into
stationarity equations \eqref{2.stat} we obtain
\begin{subequations}\label{4.stat}
\begin{equation}\label{4.xstat}
 D\yy+b_A\z=\l_1\xx,
\end{equation}
\begin{equation}\label{4.ystat}
 D\xx+ b_B\z=\l_2\yy.
\end{equation}
\end{subequations}

\subsubsection{Classification of the solutions}

Using equations \eqref{4.zero} and \eqref{4.stat} one can show that
\begin{equation}\label{4.sep}
    (D^2-\l_1\l_2)D\xx=0,\quad(D^2-\l_1\l_2)D\yy=0.
\end{equation} Consequently there exist the following three types of solutions: \begin{enumerate} \item
The first type of the solutions are related to the zero eigenvalue of $D$ and exist when both $\xx$ and
$\yy$ are the zero mode of $D$. \item The second type of the solutions are related to the middle
eigenvalue $2ab$ of $D$ and exist when \begin{equation}\label{4.clas2}
   \l_1\l_2=(2ab)^2.
\end{equation}

\item The third type of the solutions are related to the largest eigenvalue $2\mu$ of $D$ and exist when
    \begin{equation}\label{4.clas3}
   \l_1\l_2=(2\mu)^2.
\end{equation} \end{enumerate}

\subsubsection{Solutions of stationarity equations}

\paragraph{Zero mode solutions.} From Eq.\eqref{4.sep} it follows that stationarity equations
\eqref{4.stat} have the solutions $D\xx=D\yy=0$. There are four these type of solutions:
$\xx=\pm\z,\l_1=\pm b_A,\yy=\pm\z,\l_2=\pm b_B$. In what follows we will omit all the solutions with
negative Lagrange multipliers since only the solutions with $\l_1>0$ and $\l_2>0$ give a true local
maximum. Then the zero mode solution of interest is \begin{equation}\label{4.ans1}
  \xx=\z,\quad\yy=\z,\quad\l_1=b_A,\quad \l_2=b_B.
\end{equation} It gives the following product overlap \begin{equation}\label{4.g1}
   g_1^2=\frac{1}{4}(1+b_A+b_B).
\end{equation}

\paragraph{Nonphysical solutions.} Consider now solutions given by Eq.\eqref{4.clas2}. The scalar
projection of Eq.\eqref{4.stat} onto $\z$ gives $b_A=\l_1(\z\cdot\xx)$ and $b_B=\l_2(\z\cdot\yy)$, or
\begin{equation}\label{4.babb}
   b_Ab_B=\l_1\l_2(\z\cdot\xx)(\z\cdot\yy).
\end{equation} But $b_Ab_B>4a^2b^2=\l_1\l_2$, while $(\z\cdot\xx)(\z\cdot\yy)\leq1$. Hence there are no
physical solutions in this case. Sudbery et al. point out that some real roots of the characteristic
polynomial have no associated singular vectors and does not mean a local maximum~\cite{hyper}. This is
the case, the Bloch vectors of the local states are not real unit vectors and the maximal product
overlap defined by \eqref{4.clas2} is not supported by a product state. It should be neglected.

\paragraph{Main solution.} Consider now solutions given by Eq.\eqref{4.clas3}. We do not present the
derivation of the solutions but describe the main steps. First, one parameterizes unknown vectors as
follows: $\xx=(x_1,x_2,x_3)$ and $\yy=(y_1,y_2,y_3)$. Second, the condition $\l_1\l_2=4\mu^2$ forces $x_2=y_2=0$ and  factorizes the quartic equation into quadratic equations.

We  present only the solution with strictly positive Lagrange multipliers given by
\begin{subequations}\label{4.main} \begin{equation}\label{4.mainl1}
   \l_1=2\mu\s{\frac{b_A^2+4\mu^2}{b_B^2+4\mu^2}}=2(a^2+h^2),
\end{equation} \begin{equation}\label{4.mainl2}
   \l_2=2\mu\s{\frac{b_B^2+4\mu^2}{b_A^2+4\mu^2}}=2(b^2+h^2)
\end{equation} \end{subequations} and \begin{equation}\label{4.mainxy}
   \xx=\frac{{\b}_A}{b_A},\quad  \yy=\frac{{\b}_B}{b_B}.
\end{equation}
Now one can put \eqref{4.mainxy} into \eqref{4.stat}  and convince oneself that it is a
solution of the stationarity equations with the Lagrange multipliers \eqref{4.main}.

The solution \eqref{4.mainxy} gives the following product overlap \begin{equation}\label{4.maing}
   g_2^2=\frac{1}{2}.
\end{equation}

\subsection{The maximal product overlap.}

We computed the local maxima of the product overlap and now we would like to give a comment on specific
points of entanglement parameters. The invariant $b_C$ vanishes in the following two cases: either $h=0$
and $c^2+d^2 = a^2+b^2$, or $c=0$ and $d^2=a^2+b^2+h^2$. It is easy to verify that the invariant $t$
vanishes in both cases and ${\b}_A$ and ${\b}_B$ are the left and right zero modes of $G$, respectively.
On the other hand it is shown that all the five invariants are independent~\cite{inv} and can be varied
freely. What does this discrepancy mean?

The explanation is that it can happen at the edge of the region of possible values of the
invariants~\cite{private}. As a simple example, consider the three-dimensional set of points in the unit
ball, $x^2 + y^2 + z^2 \le 1$. The coordinates $x, y, z$ are independent and can be varied freely in the
interior of this region, yet we can prove that if $z = 1$ then $x = y = 0$. This is the case since
$b_C\geq0$. At minimal value of $b_C$ the sextic invariant vanishes and the correlation matrix acquires
a zero singular value. And it enables us to compute analytically maximums of the product overlap.

The maximal product overlap for the case $b_C=0$ but $h\neq0$ should be defined as
\begin{equation}\label{4.finboth}
    g^2=\max(g_1^2,g_2^2).
\end{equation} But $$ b_A+b_B = 2a\s{a^2+h^2}+2b\s{b^2+h^2}<$$ $$
a^2+(\s{a^2+h^2})^2+b^2+(\s{b^2+h^2})^2=1,$$ therefore $g_1^2<g_2^2$. Then
\begin{equation}\label{4.fine}
   g^2=1/2.
\end{equation} Hence we have shown that from $b_C=0$ it follows that $g^2=1/2$. Similarly, if either
$b_A=0$ or $b_B=0$, then $g^2=1/2$. The theorem is proved.

\section{Four- and higher-qubit states.}

An important problem is the generalization of the theorem to arbitrary $n$-qubit states. Unfortunately
this cannot be done since the theorem ceases to be true when $n\geq4$\cite{LC}. Below we illustrate what
is happening in this case.

Consider first generalized GHZ states which can be written
\begin{equation}\label{6.ghz}
   \ket{GHZ}=\cos\theta\ket{00\cdots0}+\sin\theta\ket{11\cdots1}
\end{equation}
in some product basis. The Bloch vectors are all equal in \eqref{6.ghz} and we denote
them by $\b$. Then $g^2(GHZ)=(1+|\b|)/2$~\cite{grov} and thus the theorem is valid. By the way, using
the Schmidt decomposition one can establish the same relation between $g$ and $\b$ for pure two-qubit
states. It means that pure two-qubit states with one single-qubit density matrix being completely mixed
form a single orbit under local transformations and owing to this the theorem is correct.

Consider now generalized W state which can be written
\begin{equation}\label{6.w}
\ket{W_n}=c_1\ket{100...0} + c_2\ket{010...0} + \cdots + c_n\ket{00...01}
\end{equation} in some product
basis. It is not a trivial task to show that a zero Bloch vector forces $g^2=1/2$ and viceversa in this
case. The proof can be found in Ref.\cite{dual} and the theorem is true for generalized W states too.

Consider now the four-qubit Dicke state given by
\begin{equation}\label{6.d}
 \ket{D}=\frac{1}{\s{6}}(\ket{0011}+\ket{0101}+\ket{0011}+\ket{1001}+\ket{1010}+\ket{1100}).
\end{equation}
All Bloch vectors of this state are zero, but $g^2=3/8\neq1/2$~\cite{wei-03}. Hence the
Dicke state is a counterexample showing that the straightforward generalization to higher-qubits is
impossible. Perhaps the reason is the following. In the case of pure four-qubit states the full ring of
local polynomial invariants is more complicated even in the presence of completely mixed
states~\cite{MG}. This means that in the best case the theorem can be somehow modified.

\section{Discussion}

We have shown that all pure three-qubit states that possess a completely mixed one-particle density
matrix have the same geometric measure of entanglement. This result raises several questions. For
instance, one may ask whether the theorem is valid for mixed states too. The answer is no. A simple
counterexample is a completely mixed three-qubit state whose Bloch vectors are all zero but the state is
separable.

The most important question is whether the theorem can be extended to other entanglement measures
suitable for arbitrary multipartite states. There are two sound arguments that the extension is indeed
possible. First, the keystone idea of the proof is that at the edge of a quartic polynomial the sextic
polynomial vanishes and the algebraically dependent invariant becomes redundant. And then two Bloch
vectors become left and right zero modes of the correlation matrix. But this is a specific feature of
local invariants of pure three-qubits and, therefore, is related directly to quantum entanglement. This
feature has no relation to the geometric measure, or, more precisely, it is the same peculiarity for all
entanglement measures. Hence any reliable entanglement measure should detect this peculiarity as a
specific point. Second, as it is pointed out in Ref.\cite{shared}, all of the states with $g^2=1/2$,
designated as shared quantum states,  can be used as a quantum channel for the perfect teleportation and
dense coding. Therefore they must possess the same amount of entanglement and thus a reliable
entanglement measure should not vary on the manifold of shared quantum states. Again one can conclude
that a completely mixed state should be a critical point for a good multipartite measure.

Another important question is: what is happening with general n-qubit states? We have seen that both
two- and three-qubit pure states have the same type of critical points that are the edge values of Bloch
vectors. And when $n\geq4$ the theorem ceases to be true. But our brief analyze in Sec. VI shows that
the situation is more complicated. Indeed, for GHZ- and W-class states the theorem works well while it
is wrong for Dicke states. This may indicate that there are different types of critical points at $n=4$.
And GHZ and W states have one type of critical points and Dicke states have another types of a such
points. Then we would like to know how many types of critical points exist in the space of entanglement
parameters of pure four-qubit states. And what is the role of those points or what kind of quantum
phenomena are behind them. To clarify these points we need to analyze carefully the complete set of
polynomial invariants of pure four-qubit~\cite{MG} and two-qubit mixed~\cite{makh} states.

Finally we would like to discuss whether the inverse theorem is true. In the case of pure two-qubit
states it is an easy task to show that if $g^2=1/2$ then $b_A=b_B=0$. But in the case of pure
three-qubit states the problem is open. The question is: does from $g^2=1/2$ it follow that
$b_Ab_Bb_C=0$? Unfortunately we failed to prove or disprove the inverse theorem.

\begin{acknowledgments}
We thank Markus Grassl for useful comments. This work was supported by ANSEF Grant No. PS-1852.
\end{acknowledgments}

\end{document}